\documentclass[10pt, conference]{IEEEtran}
\usepackage{makecell}
\usepackage{cite}
\usepackage{color}
\usepackage{colortbl}
\usepackage{array}
\usepackage{graphicx}
\usepackage{multirow}
\usepackage{ragged2e}
\usepackage{amsmath}
\usepackage[export]{adjustbox}
\IEEEoverridecommandlockouts
\usepackage{comment}
\usepackage{pifont}
\usepackage{todonotes}
\usepackage{subcaption}
\usepackage{listings}
\usepackage{flushend}
\usepackage{tikz}
\usepackage{textcomp}
\usepackage[doipre={DOI:~}]{uri}
\usepackage{lipsum}
\newcommand\copyrighttext{%
  \footnotesize \textcopyright 2022 IEEE. Personal use of this material is permitted.  Permission from IEEE must be obtained for all other uses, in any current or future media, including reprinting/republishing this material for advertising or promotional purposes, creating new collective works, for resale or redistribution to servers or lists, or reuse of any copyrighted component of this work in other works.
 
  Published as a conference paper at the 2023 IEEE/ACM International Symposium on Low Power Electronics and Design (ISLPED).
  ~\doi{10.1109/ISLPED58423.2023.10244587} }
\newcommand{\copyrightnotice}{
\begin{tikzpicture}[remember picture,overlay]
\node[anchor=south,yshift=10pt] at (current page.south) {\fbox{\parbox{\dimexpr\textwidth-\fboxsep-\fboxrule\relax}{\copyrighttext}}};
\end{tikzpicture}%
}

\title{Model-Driven Dataset Generation \\for Data-Driven Battery SOH Models}
\author{
    \IEEEauthorblockN{Khaled Sidahmed Sidahmed Alamin\IEEEauthorrefmark{1}, Francesco Daghero\IEEEauthorrefmark{1},
    Giovanni Pollo\IEEEauthorrefmark{1},\\
    Daniele Jahier Pagliari\IEEEauthorrefmark{1}, Yukai Chen\IEEEauthorrefmark{2},
    Enrico Macii\IEEEauthorrefmark{1}, Massimo Poncino\IEEEauthorrefmark{1}, Sara Vinco\IEEEauthorrefmark{1}}
    \IEEEauthorblockA{\IEEEauthorrefmark{1}Politecnico Di Torino, Turin, Italy \{name.surname@polito.it\}
    \IEEEauthorrefmark{2}IMEC, Leuven, Belgium \{yukai.chen@imec.be\}}
}

\begin{document}

\bstctlcite{IEEEexample:BSTcontrol}

\maketitle
\copyrightnotice

\begin{abstract}
    Estimating the State of Health (SOH) of batteries is crucial for ensuring the reliable operation of battery systems. Since there is no practical way to instantaneously measure it at run time, a model is required for its estimation.
Recently, several data-driven SOH models have been proposed, whose accuracy heavily relies on the quality of the datasets used for their training. Since these datasets are obtained from measurements, they are limited in the variety of the charge/discharge profiles. 

To address this scarcity issue, we propose generating datasets by simulating a traditional battery model (e.g., a circuit-equivalent one). The primary advantage of this approach is the ability to use a simulatable battery model to evaluate a potentially infinite number of workload profiles for training the data-driven model. Furthermore, this general concept can be applied using any simulatable battery model, providing a fine spectrum of accuracy/complexity tradeoffs. Our results indicate that using simulated data achieves reasonable accuracy in SOH estimation, with a 7.2\% error relative to the simulated model, in exchange for a 27X memory reduction and a $\approx$2000X speedup.
\end{abstract}


\section{Introduction}
\label{sec:intro}
\everypar{\looseness=-1}
The accuracy of onboard State of Health (SOH) estimation in Battery Management Systems (BMS) is essential for ensuring the safety and reliability of battery systems of a battery-powered device, and in particular for Electric Vehicles (EVs).
As there is no practical physical way to instantaneously measure the SOH, such tracking inevitably requires a model.
The literature about SOH models is extremely vast, including
electrochemical, equivalent circuits,  semi-empirical, analytical, and statistical models \cite{su131810042}.
More recently, on the wave of the Machine Learning (ML) hype, a new category of {\em data-driven models} has emerged, in which a set of instantaneously measurable battery parameters (typically, voltage, current, and temperature) relative to a charge or discharge session of a battery is labeled with corresponding SOH values calculated at session's end \cite{9454160}.
These labeled measures are then used as a dataset to train appropriate ML models
\cite{9036949,Heinrich2021}.
Data-driven models essentially solve the two main drawbacks of traditional models: 
(i) they are more general, as models for different battery types can be naturally obtained by training them with measurements on different devices, thus also covering  variability aspects; and (ii) they do not require any kind of simulation, thus 
resulting in significant reductions in time and space complexity when deploying the model on resource-constrained devices (i.e., no need for a large amount of simulation operations to estimate battery dynamics and/or for a simulation engine).

On the other hand, the quality of data-driven models is strongly dependent on the size and the variety of the dataset.
As these datasets are obtained by experimental measurements, it is \textit{materially unfeasible to provide an acceptable coverage of the design space}: datasets are generated at specific working conditions, determined by the application domain, and in limited time, thus restricting the variety of explored charge/discharge/rest patterns, discharge current profiles, and load currents. 
Last but not least, such a large exploration space should possibly be repeated on multiple battery instances in order to account for the intrinsic variability of the devices. The consequence is that datasets obtained by measurements are by definition very accurate, but accuracy is guaranteed only in the few points of the experiment space that have been measured. \everypar{\looseness=-1}

The key motivation of this paper is to fundamentally swap this asymmetry, i.e.,  to \textit{sacrifice some accuracy while extending the coverage of the design space}.
We propose \textit{to use measurements} (possibly much fewer than those required to generate the whole dataset) \textit{to build a simulatable battery model} that incorporates the desired effects. This model can then be used to generate arbitrarily large datasets, which will serve as the training set for constructing more lightweight and flexible data-driven models.

Several are the advantages of this approach:
\begin{itemize}
    \item \textbf{Exploration of virtually unlimited data points:} once the battery model is built, any kind of current and/or temperature workload can be simulated to generate as many data points as needed, at a lower cost and in less time than would be required for experimental measurement setups;
    \item \textbf{Tunable accuracy/complexity tradeoff:} depending on the quality of the available measurement data, more or less accurate battery models can be built, providing a flexible range of datasets for the data-driven model;
    \item \textbf{Possibility of model combination:} multiple types of battery models can be built \cite{su131810042}) 
    and integrated to cover different aspects of battery dynamics, increasing the comprehensiveness of the final battery model;
    \item \textbf{The final SOH model is still a data-driven SOH model:} its execution does not require simulation and it is essentially a callable function of ``live'' parameters that can be deployed on a target resource-constrained device.
\end{itemize}
Our results show that using data obtained from the chosen simulation models~\cite{simulinkModel}, we achieved an error of about 7\% with respect to the SOH data while reducing memory usage by 27X and speeding up calculations by~$\approx$2000X.

\section{Background and Related Work}
\subsection{Battery Aging}
\label{subsec:aging}
Battery aging is the effect of %
 (i) \emph{calendar aging} ($L_{cal}$), reflecting  battery intrinsic degradation when in rest conditions as an effect of temperature, State of Charge (SOC), and elapsed time; and (ii) \emph{cycle aging}, representing capacity loss during each charge/discharge cycle ($L_{cyc}$), depending on average values of current $I$, SOC, cell temperature $T$ and Depth of Discharge (DoD, i.e., difference between final and initial SOC) \cite{5619782,7357135,VETTER2005269}.
Overall capacity loss $(L_C)$ is thus the sum of a global term for calendar aging plus the sum of the degradation in each cycle \cite{7488267}:
\begin{multline}
    L_C (t,SOC,DoD,I,T) = \\ L_{cal}(t,SOC,T) + \sum_{i=1}^{N} L_{cyc} (I_i,SOC_i,DoD_i,T_i)
    \label{eq:caploss}
\end{multline}
where $N$ represents the number of charge/discharge cycles, SOC and $T$ are average over an interval of length $t$ in $L_{cal}$, and refer to each individual cycle $i$ in $L_{cyc}$. 
State-of-the-art models for $L_{cal}$ and $L_{cyc}$ leverage either the similarities of fatigue process of materials subjected to cyclic loading \cite{5619782} or incorporate electro-chemical properties of the charge/discharge process \cite{7488267}. 
It is outside the objective of this work to provide further details about the models themselves; an exhaustive overview of these models is available in \cite{VENNAM2022104720}.

\subsection{Data-Driven SOH Estimation}
The relative simplicity of casting the estimation of the SOH as the problem of building a predictive model has spurred a number of datasets available online \cite{dataset1} and a vast literature about models\cite{9036949}.  

\everypar{\looseness=-1}The approaches used to estimate SOH differ in two key aspects: how SOH is measured and the ML model used for estimation.

Concerning the former aspect, SOH is measured either in terms of the loss of capacity (which is more typical) or in terms of the increase of the internal resistance.
For the latter aspect, conversely, the spectrum of options is definitely much wider: models range from various types of Neural Networks (NNs, feed-forward or recurrent ones) to simpler models like random forests, Support Vector Machines (SVMs), or Bayesian networks.
Many of these approaches claim to estimate capacity or resistance with high accuracy, making them promising candidates for SOH estimation.

However, as emphasized by the authors of \cite{9036949}, comparing different approaches and establishing reference models is challenging. One reason is the \textit{quality} of the datasets. 
As a matter of fact, most of these datasets are too limited in size. 
Besides the obvious impact that small datasets have on the accuracy of data-driven models (in particular NNs), there is also the problem of the \textit{variety} of the dataset points. As they are obtained from lab measurements, there are some intrinsic limitations in generating some specific data points (e.g., very low load currents, which will require prohibitive runtimes) and are susceptible to measurement errors and noise.

\section{Proposed Methodology}
\subsection{Workflow of the Proposed Methodology}
Our idea is to use the datasets mentioned in the former section (or possibly a small portion thereof) to build a full-fledged, simulatable battery model including the SOH \textit{together with the entire battery dynamics}, and then use this simulatable model to generate additional data points, that become the training set for a higher quality SOH data-driven model.

\begin{figure}[!htbp]
	\begin{center}
		\includegraphics[width=.95\linewidth]{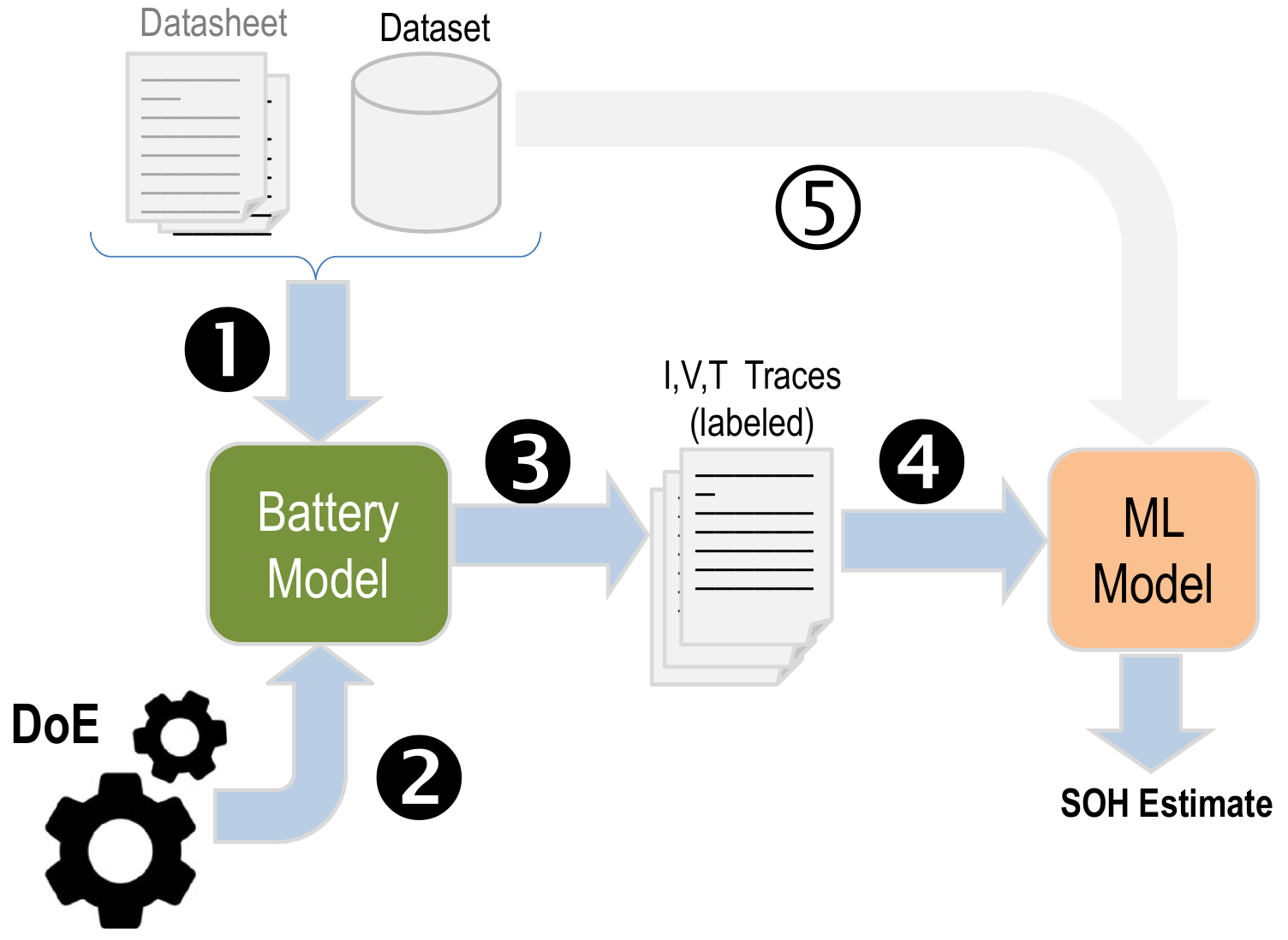}
		\caption{Conceptual flow of the proposed methodology}
		\label{fig:flow} 
	\end{center}
\end{figure}

Figure~\ref{fig:flow} sketches the envisioned flow to implement this approach.

The flow starts (\ding{202}) with an \emph{available set of battery data}, which typically includes measurements collected at various operating conditions and can be generated afresh or obtained from public datasets. While datasheets can be used to obtain battery information, they generally result in poorer accuracy in the construction of the model~\cite{petricca}. 

Battery datasets are then used to \emph{identify the parameters of a battery model} (hereafter the \textit{simulation model}), which tracks (at least) the desired target quantity (in our case, SOH). Depending on the type of model, an appropriate procedure is followed for the identification of the model parameters \cite{petricca,barreras,bocca}. Section \ref{sec:interface} elaborates on the requirements for the simulation model, and Section \ref{sec:models:simulation} surveys the main ones available in the literature that comply with these requirements.

Once the simulation model is built, {\em an exhaustive set of synthetic traces is generated} in a Design-of-Experiment (DoE) step (\ding{203}), exercising as many points as possible in the space of the model inputs. For each of these design points, a simulation of the battery model is run to yield one output trace (\ding{204}).

Finally each trace generated by the simulation model is used for training the SOH ML model (hereafter the \textit{data-driven model}, \ding{205}). Section \ref{sec:models:datadriven} will describe the various options for the data-driven models and their impact on a BMS. 
If the format of the traces \ding{202} and \ding{204} are the same, we can use a mix of real (measured) and simulated (model-driven) traces to train or test the model (\ding{196}). This step is not explicitly depicted in the flow, but it showcases the flexibility of the approach.

\subsection{General Requirements for Battery Simulation Models}\label{sec:interface}
\begin{figure}[!htbp]
	\begin{center}
		\includegraphics[width=0.8\linewidth]{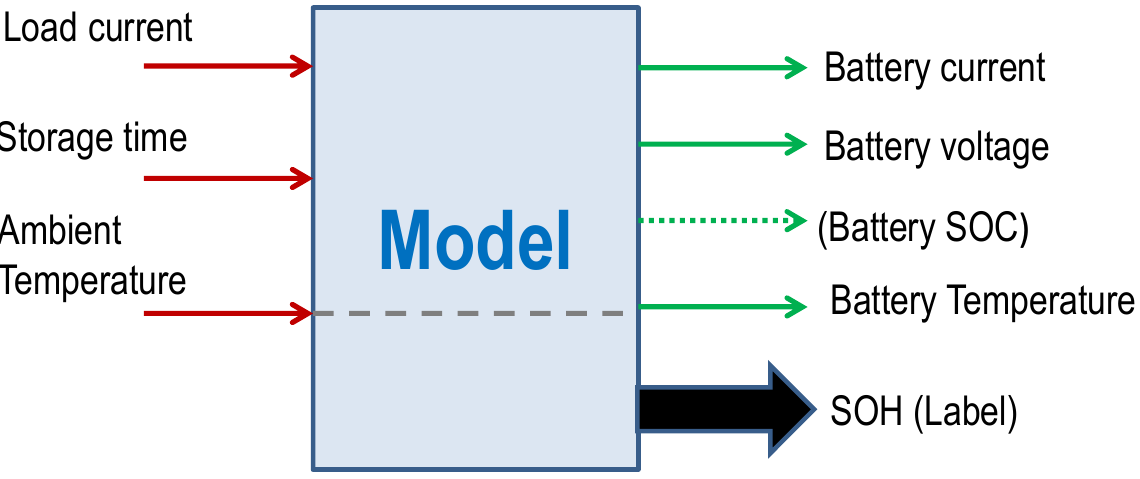}
		\caption{General interface of the battery simulation model.}
		\label{fig:model} 
	\end{center}
\vspace{-0.5cm}
\end{figure}

The need for a simulatable model raises the issue of identifying a common  \textit{interface of the model}, defining the requirements in terms of modeled quantities. The general interface used in our framework is shown in Figure~\ref{fig:model}.

The inputs of the models are the independent variables that represent the external conditions under which the battery is used, and must be measurable at runtime.
Inputs are load (charge/discharge) current, ambient temperature, and total storage time (needed for calendar aging). Notice that although the general aging model (Equation \ref{eq:caploss}) contains the number of cycles $N$, this is not strictly required as an input of the battery model; $N$ can in fact be simply obtained from the dataset by using the timestamp of each data point: a new cycle is counted anytime we complete a charge/discharge sequence.

The outputs of the model are typically arbitrary and can include a range of quantities typical of most public datasets \cite{dataset1}. However, two constraints must be considered. Firstly, traces generated by the battery model in \ding{204} \textbf{must be labeled}, meaning that they must include a label that represents SOH, so that the traces can be used to train the ML battery model effectively. Secondly, if possible, the simulation model traces should \textit{follow the same format of the datasets used in \ding{202}}. This will allow the synthetic dataset to be used as an extension of the original dataset for training and/or testing the resulting data-driven battery model, thereby enabling a wider scenario. 
Finally, it is worth emphasizing the importance of temperature in modeling SOH and highlighting the distinction between \textit{ambient} and \textit{battery} temperature. If the model is chosen carefully, it will include a thermal model that, based on the electrical quantities \textbf{and} the external temperature, can accurately predict the battery's internal temperature. However, if the thermal model is not available, a rough approximation is to use the ambient temperature as a proxy for the internal temperature (the dashed line). This is essential because SOH is highly sensitive to temperature.

\subsection{Choice of the Battery Simulation Model}\label{sec:models:simulation}
The requirements implied by the model interface defined in the previous section might result in a possible difficulty in our approach.
Fortunately, models with these characteristics are available in the literature \cite{7281923}, with different ranges of accuracy and complexity. Electrochemical models represent the chemical reactions as differential equations \cite{5256311}, but they are too complex to be executed on board in real-time. Circuit-equivalent models like \cite{5212025,7420729} require an RC network solver or a state-space computation environment and may result in being very heavy; improving their performance on the other hand implies a reduction in accuracy. Kalman filters are a good compromise: they allow to specify an error bound and, being an iterative process, they tend to reduce the estimation error to zero \cite{5289654}.

\begin{figure}[!htbp]
	\begin{center}
		\includegraphics[width=\linewidth]{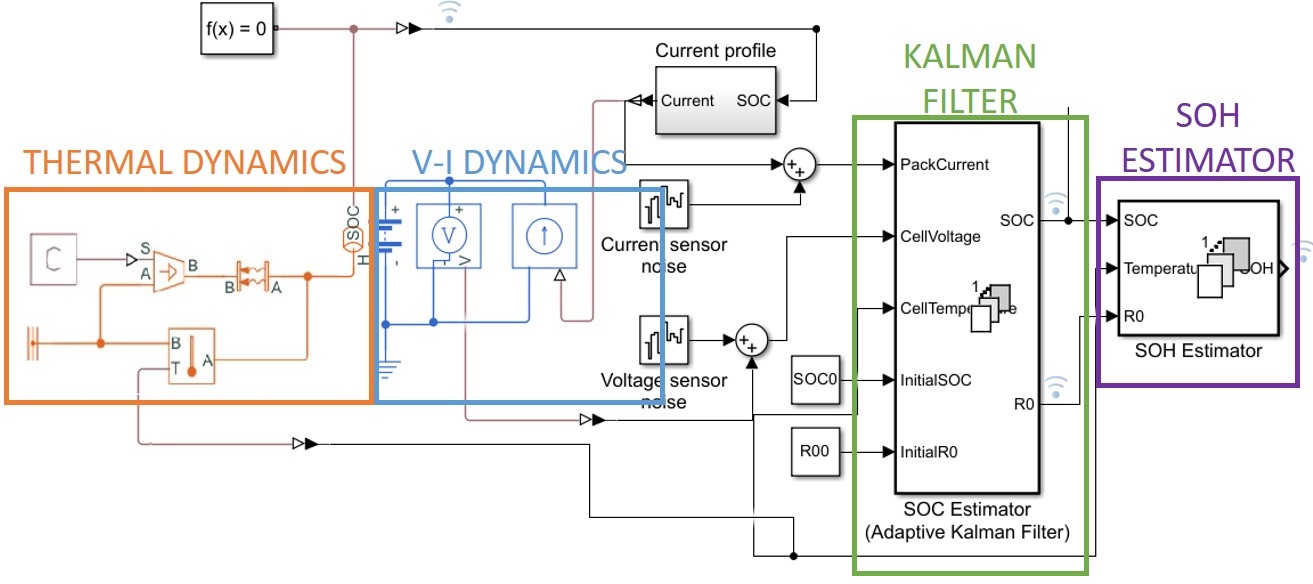}
		\caption{Adopted battery model based on Kalman filters.}
		\label{fig:kalman} 
	\end{center}
  \vspace{-0.5cm}
\end{figure}

For the aforementioned reasons, we chose the Kalman filter-based simulation model of \cite{simulinkModel} as a reference for this work (Figure \ref{fig:kalman}). This pre-defined model is designed to describe a 27Ah battery and uses Simscape to represent both the thermal and electrical dynamics of the battery. 
The model reproduces capacity fading due to thermal cycling and uses an unscented Kalman filter to estimate battery SOC and internal resistance based on values of voltage, current and internal temperature at runtime. These estimates are then used as inputs of a SOH estimation block built with measurement-based lookup tables. 

\subsection{Choice of the Battery Data-Driven Model}\label{sec:models:datadriven}
The last step of the flow is the training of a data-driven model on the dataset generated by the simulatable model. 
The proposed flow is \emph{independent of the target data-driven model}: in principle, any type of regressor can be embedded in the ``ML Model'' block of Figure~\ref{fig:flow}, ranging from a simple linear model to a complex deep NNs: this %
offers a much finer-grain tradeoff between accuracy and computational complexity than traditional physics-based models.

We focus on a scenario in which SOH estimation must be implemented on a low-power Microcontroller (MCU), possibly hosted in the BMS, to enable onboard SOH estimation. To achieve this, we selected two lightweight data-driven models: (i) a Gradient Boosted Tree (GBT) regressor, i.e., an ensemble of decision trees selected for its very fast prediction consisting of a small number of branching operations~\cite{daghero}, and (ii) a Multi-Layer Perceptron (MLP), i.e., a simple fully-connected neural network appropriate for dealing with pre-aggregated features. 
These models are just examples to prove the flexibility of our proposed approach, and the most suitable data-driven model should be selected based on the desired SOH estimation accuracy and the constraints of the target platform.

\subsubsection{Model Training}
The training dataset is populated from the raw samples of battery current ($I$), voltage ($V$), and temperature ($T$) generated by the simulation model at the frequency $f_s$ that we expect onboard measurements to occur ($f_s =$ 1 Hz in our experiments, but this highly depends on the time constants of the considered system).
The ML regressors are then trained on 12 statistical features of $I$, $V$ and $T$, aggregated over a \textit{time-window} of configurable duration $W$ (2 hours in our experiments): 
\textit{mean}, \textit{variance}, \textit{min}, and \textit{max} of each quantity. %
Given that we target deployment on a highly-constrained, low-performance MCU, we avoid more complex features (e.g., Skewness or Kurtosis)

since the computational cost for their calculation could outweigh the resulting accuracy improvement. Given that windows have a constant duration, it is not necessary to include also the elapsed time. %

ML models are trained to predict $\Delta SOH = SOH_{initial} - SOH_{final}$ in each window,

normalized to [0:1] for numerical stability. The training loss function for both models is the Mean Squared Error (MSE) between their prediction and the $\Delta SOH$ estimated by the simulation model. Note that the raw SOH prediction can be recovered by de-normalizing and accumulating predictions over consecutive windows (1 multiplication and 1 sum per window).

\subsubsection{Model Space Exploration} 
A large number of design points can be obtained by (1) exploring model hyper-parameters, and (2) selecting model features. 
Concerning hyperparameters exploration, we vary the number of estimators (decision trees) of the GBT and their maximum depths in the sets [5, 10, 20, 50, 100, 200] and [1, 2, 3, 4, 5, 10, 30, 50] respectively, for a total of 48 configurations. Similarly, we consider 42 MLP variants, with 1 to 2 hidden layers of sizes in [4, 8, 16, 32, 64, 128].
We explore these options with grid search and 50/20/30\% train/validation/test split to obtain a Pareto-frontier in the MSE vs. execution time and MSE vs. memory occupation spaces. For the MLP, we use the Adam optimizer with a batch size of 64 and a learning rate of 0.001, training for 50 epochs. The remaining hyper-parameters are kept at the default values of the respective training libraries (see Section~\ref{sec:res_setup}).

A selection of the best feature set, down to a minimum of 3 features, is applied by using a Recursive Feature Elimination (RFE) algorithm. 
Thus, the grid search described above is repeated 10 times, once for each feature sub-set, resulting in a total of 480 GBT and 420 MLP models being evaluated.

Since the deployment of such a large number of model variants would be impractical, we use simple mathematical models in the exploration phase to estimate the time/energy and memory complexity of each point. 

For time and energy estimation, we count the operations required for the two steps involved when using the model:
\begin{itemize}
    \item \textbf{Feature Extraction}: 
Since all considered features can be extracted with $O(N_{samples}$) operations, where $N_{samples} = \frac{W}{f_s}$ is the number of samples in a window, we estimate feature extraction time as $\frac{W}{f_s} \cdot N_f$, where $N_f$ is the number of features.
    \item \textbf{Model Evaluation}:
MLP evaluation cost is obtained by counting the total number of multiply-and-accumulate (MAC) operations, while GBT cost is obtained by counting the number of branch operations.
\end{itemize}
Since both phases consist mainly of arithmetic operations (with no division), we use the above two time quantities also as proxies of energy consumption: this is reasonable as the exploration relies on relative comparisons and it is architecture-independent, so we can reasonably assume that the energy cost is roughly equivalent to the execution time times the average power cost of an arithmetic operation.

With these models, we note that feature extraction time is one or two orders of magnitudes higher than model evaluation time, especially for GBTs.

Concerning memory, we estimate the cost of an MLP configuration as the number of bytes required to store the network weights and the two largest activation buffers~\cite{pulp_nn}.
For the GBT, we count the bytes required to store the ensemble data structure and the input feature buffer~\cite{daghero}.

While simplified, these models are effective in preserving complexity rankings among different configurations, especially for simple hardware like an MCU. On the other hand, using them allows us to perform the entire training and hyper-parameters search process in less than 4 hours on a laptop.

\section{Experiment Results}\label{sec:results}
\begin{figure*}[ht]
  \centering
  \begin{subfigure}[t]{.45\textwidth}
      \includegraphics[width=\textwidth]{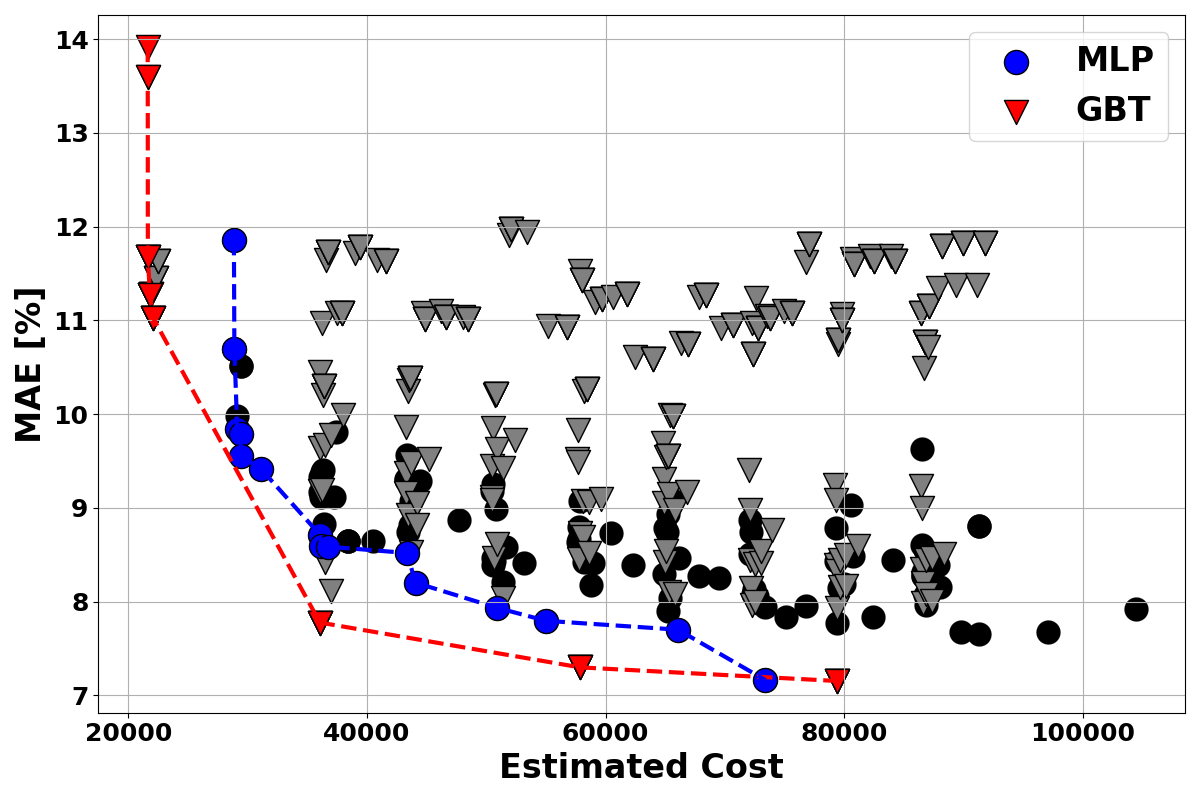}
      \caption{Time/Energy}
  \end{subfigure}
  \begin{subfigure}[t]{.45\textwidth}
      \includegraphics[width=\textwidth]{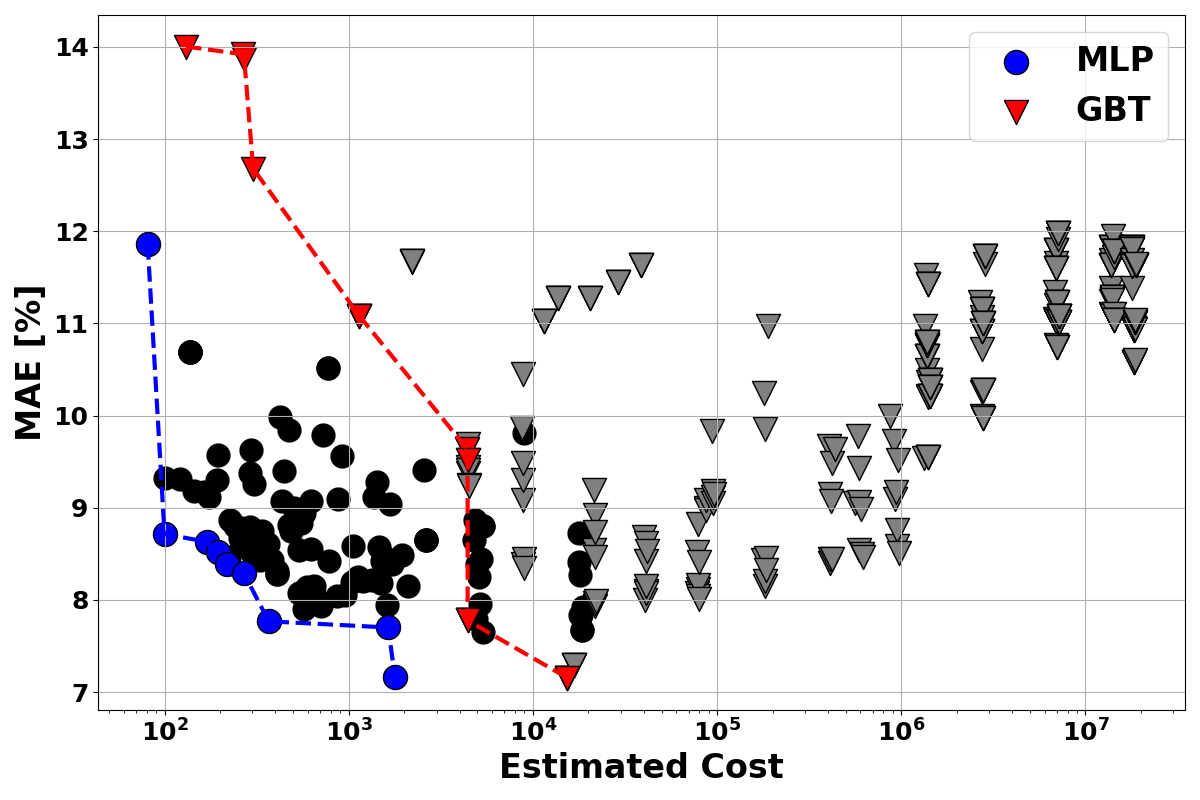}
      \caption{Memory}
  \end{subfigure}
  \caption{Pareto-fronts obtained from the hyper-parameters exploration of data-driven models}
  \label{fig:fronts}
  \vspace{-0.4cm}
\end{figure*}

\subsection{Experiment Setup}\label{sec:res_setup}
We train data-driven models using the Scikit-Learn and Keras Python libraries for GBT and MLP respectively. As the target embedded device, we consider the ultra-low-power RISC-V MCU PULPissimo~\cite{quentin}, onto which we deploy both GBTs and MLPs using optimized  libraries written in C. For GBTs, we leverage an in-house implementation similar to the one described in~\cite{daghero} for random forests, whereas for the MLPs we use a single-core version of the PULP-NN library~\cite{pulp_nn}. Time and energy results refer to a 22nm realization of PULPissimo working at 205.1 MHz~\cite{quentin}.

\subsection{Dataset Generation}\label{sec:dataset}
Simulation data is generated by configuring the model of~\cite{simulinkModel} with the default parameters. We run a set of simulations, each lasting a maximum of 1,000 hours, or until battery SOH reaches 0. Each simulation uses either a different temperature or a different current profile. Specifically, we consider ambient temperature values in the [10$^{\circ}$C : 40$^{\circ}$C] range with a step of 5$^{\circ}$C. We stimulate the battery model with constant, square-wave, and ``random walk'' load current profiles for both charge and discharge cycles, with values ranging from $\pm$0.25A to $\pm$2A. Each current pulse in a random walk has a duration of $\approx$1 minute, whereas square waves have a period of $\approx$30 minutes. Combining these conditions, we obtained 110 simulations that, once aggregated in non-overlapping windows of length $W = 2$ hours, gave us 17,842 samples. 

We split those data into training, validation, and test sets with 50/20/30\% proportions for all our experiments. Importantly, the split is performed \textit{at the simulation level} (i.e., not at the window level), meaning that windows belonging to the same simulation \textit{cannot} be simultaneously present, for instance, in the training and test sets. This is the most realistic scenario, since in order to perform well, the data-driven models must learn to extrapolate the $\Delta$SOH for different simulation conditions w.r.t. those seen during training.

Notably, this setup would allow us to easily generate more data, for example conducting an error analysis to identify the $T$ and $I$ conditions in which our GBT/MLP models perform worse and enhancing the dataset accordingly.

\subsection{Pareto Analysis}
Figure~\ref{fig:fronts} shows the results of model space exploration for data-driven models. The x-axes of the two plots report the time/energy and memory estimated cost respectively, according to the models of Sec.~\ref{sec:models:datadriven}, whereas the y-axis reports the $\Delta$SOH MAE with respect to the simulation model, in percentage. In both charts, each dot/triangle refers to one MLP/GBT hyper-parameter and input features configuration, and the blue/red points highlight the respective Pareto fronts.

Tuning models configurations, we obtain Pareto-optimal solutions spanning a 4x range in estimated time/energy and more than 2 order of magnitudes in memory occupation, with MAEs ranging between 7\% and 14\%. GBT models achieve superior results in terms of error vs. time/energy trade-off, but have higher estimated memory than MLPs. This demonstrates the flexibility achievable by selecting from the rich spectrum of data-driven model architectures.

Table~\ref{tab:pareto} reports the detailed results of the \textit{extremes} of the two Pareto curves. Namely, for each model type, we report the configuration achieving the lowest error (-E suffix), the lowest estimated time (-T), and the lowest estimated memory (-M). Note that the latter two coincide with the MLP. Besides the precise number of features, the hyper-parameter setting, and the MAE, we also report two additional error metrics, i.e., the Mean Squared Error (MSE) and the R$^2$ score. These results show that both hyper-parameter tuning and feature selection are important to find optimal data-driven model configurations.

\begin{table}[ht]
\caption{Extremes of the Pareto-curve.}
\label{tab:pareto}
\resizebox{\columnwidth}{!}{
\begin{tabular}{l|l|l|l|l|l|l}
\multicolumn{7}{c}{\textbf{GBT}} \\\hline
Model & \# Feat. & \# Trees & Max Depth & MAE [\%] & MSE [\%] & R$^2$\\\hline
GBT-E & 11 & 50 & 5 & 7.15 & 0.96 & 0.729\\
GBT-T & 3  & 5 & 2 & 13.92 & 2.87 & 0.182\\
GBT-M & 4  & 5 & 1 & 14.00 & 3.14 & 0.107\\
\hline
\multicolumn{7}{c}{\textbf{MLP}} \\\hline
Model & \# Feat. & \# Layers & Hidden Size & MAE [\%] & MSE [\%] & R$^2$\\\hline
MLP-E & 10 & 3 & 128   & 7.16  & 0.93 & 0.736\\
MLP-T & 4  & 3 & 8 & 11.86 & 2.04 & 0.420\\
MLP-M & 4  & 3 & 8 & 11.86 & 2.04 & 0.420\\
\hline
\end{tabular}
}
\vspace{-0.25cm}
\end{table}

\subsection{Deployment Results}

\begin{figure*}[ht]
  \centering
  \includegraphics[width=0.4\textwidth]{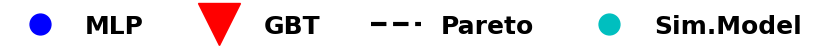}
  
  \begin{subfigure}[t]{.49\textwidth}
      \centering
      \includegraphics[width=0.8\textwidth]{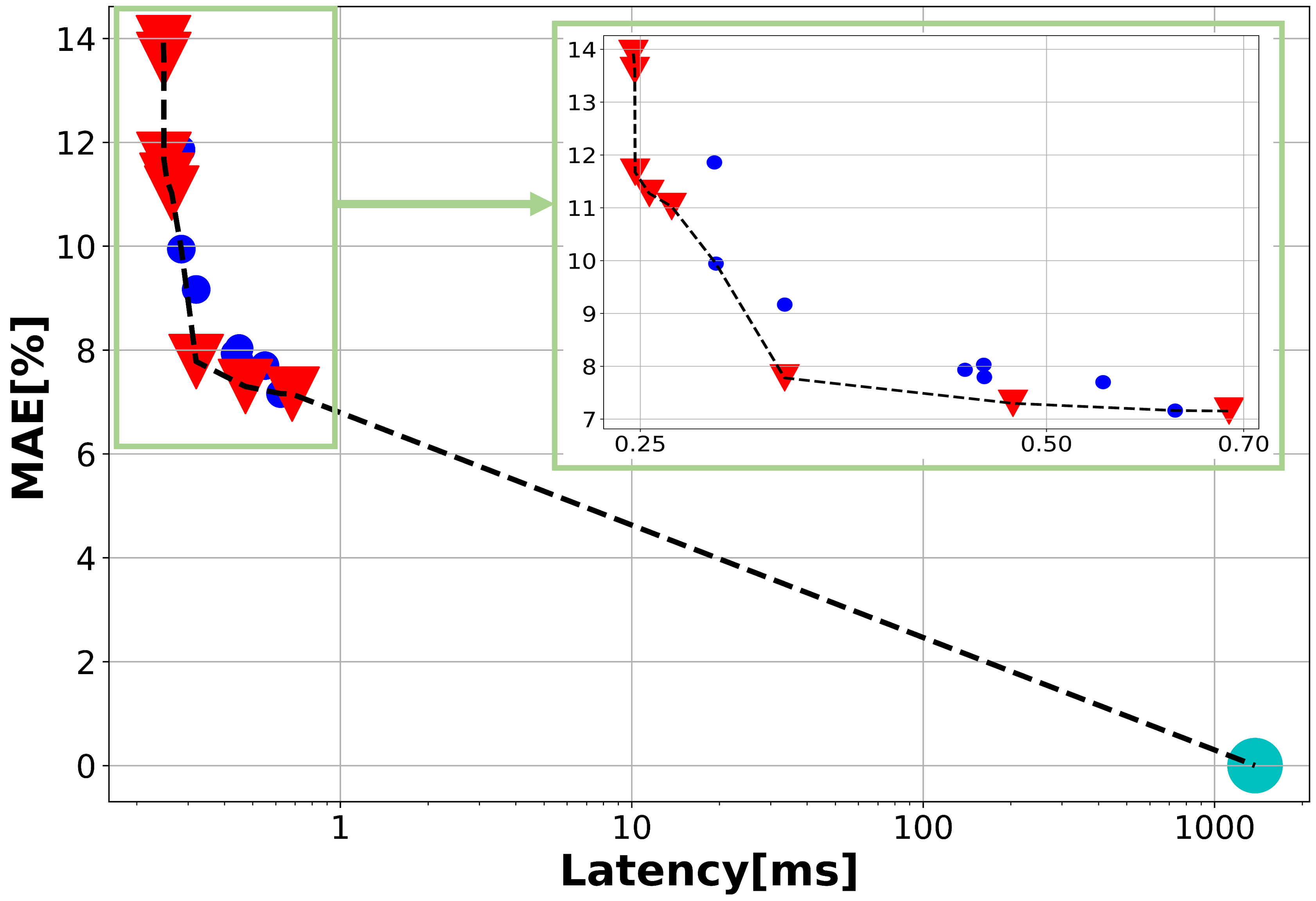}
      \caption{Time}
  \end{subfigure}
  \begin{subfigure}[t]{.49\textwidth}
      \centering
      \includegraphics[width=0.8\textwidth]{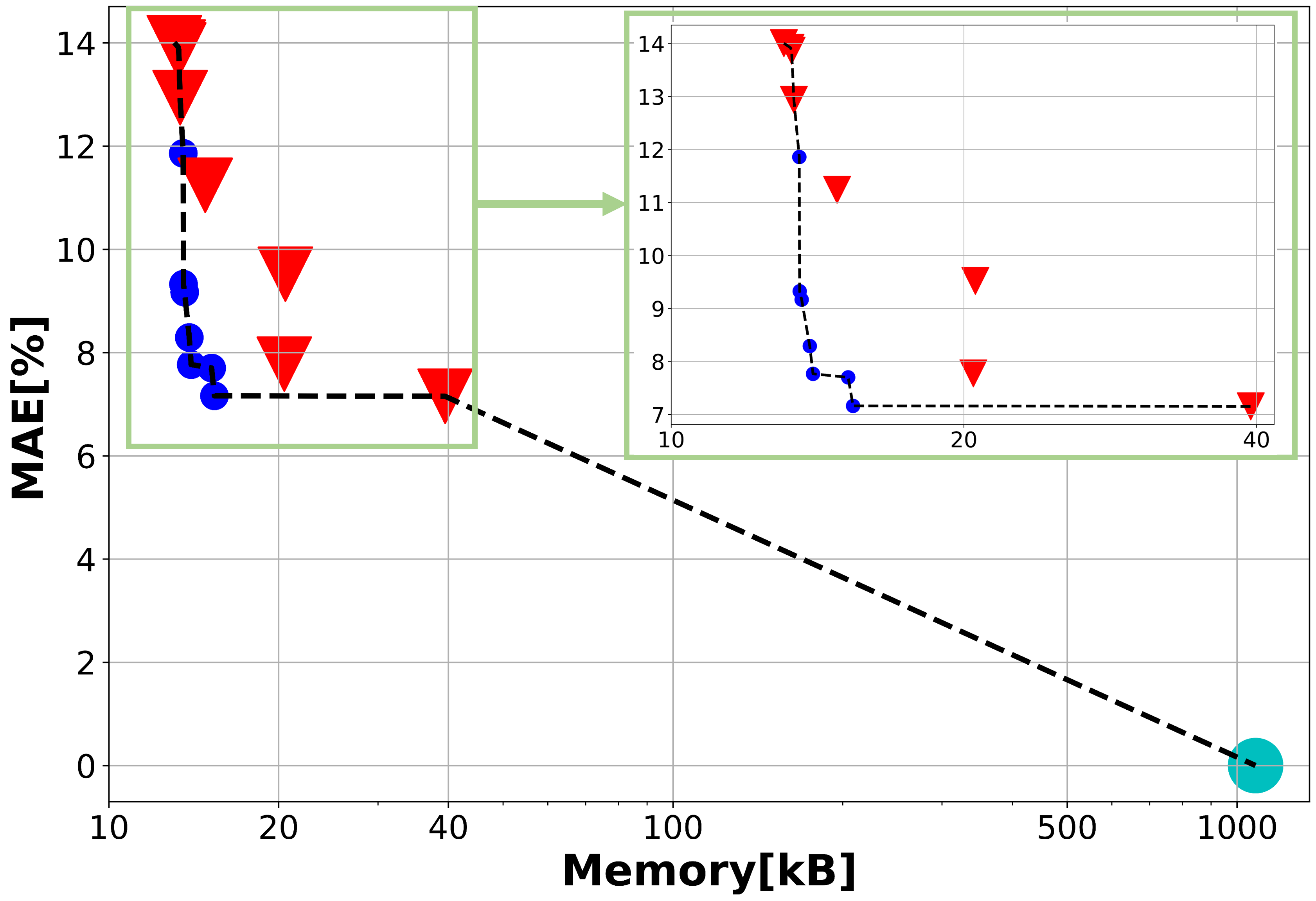}
      \caption{Memory}
  \end{subfigure}
  \caption{Deployed data-driven models and comparison with the simulation-based model.}
  \label{fig:deployment}
  \vspace{-0.4cm}
\end{figure*}

Figure~\ref{fig:deployment} reports the results of deploying all Pareto-optimal models from Figure~\ref{fig:fronts} on PULPissimo, thus replacing the cost estimates with actual latency and memory (data plus code) measures on the target. %
The detailed deployment results for the extremes of the MLP and GBT Pareto fronts (same models of Table~\ref{tab:pareto}) are also reported in the first six rows of Table~\ref{tab:deployment}, in terms of the number of clock cycles, latency, total memory occupation, and energy consumption per regression.

In order to compare the time and memory costs of our data-driven models against those incurred by deploying a simulation-based model on-device, we compiled our ground truth reference from~\cite{simulinkModel} to a binary executable through the Simulink coder toolbox. We targeted a laptop-class CPU (Apple M1 Pro), since~\cite{simulinkModel} turned out to be impossible to compile for our RISC-V embedded target, as the Simulink coder relies on pre-compiled, x86-only support libraries, and the memory required by this model  exceeds the 512kB L2 available on the target.

To perform a fair comparison, we also compiled our lowest-MAE data-driven model (GBT-E) for the M1 Pro, using the C-based library of~\cite{daghero}. The results are presented in the two rows marked with $\dagger$ in Table~\ref{tab:deployment}. The last row estimates the figures of the simulation model on PULPissimo by re-scaling the simulation model results on M1 Pro to the RISC-V, using the ratio of the results obtained by GBT-E on the two platforms as proportionality factors. The corresponding latency and memory values are also represented as light-blue dots in Figure~\ref{fig:deployment}.

\begin{table}[ht]
\caption{Deployment Results.}
\label{tab:deployment}
\resizebox{\columnwidth}{!}{
\begin{tabular}{l|l|l|l|l|l}
Model & MAE [\%] & Cycles & Latency [ms] & Memory [kB] & Energy [$\mu J$] \\\hline
GBT-E & 7.15 & 140$\cdot10^3$ & 0.68 & 39.46 & 2.6 \\
GBT-T & 13.92 & 51$\cdot10^3$ & 0.25 & 13.26 & 0.94\\
GBT-M & 14.00 & 58$\cdot10^3$ & 0.28 & 13.06 & 1.04\\
\hline
\hline
MLP-E & 7.16 & 122$\cdot10^3$ & 0.6 & 15.38 & 2.36\\
MLP-T & 11.86 & 58$\cdot10^3$ & 0.28 & 13.54 & 1.01 \\
MLP-M & 11.86 & 58$\cdot10^3$ & 0.28 & 13.54 & 1.01 \\
\hline
\multicolumn{6}{c}{\textbf{Simulation-model Comparison}} \\\hline
GBT-E$^\dagger$ & 7.2 & 560$\cdot10^3$ & 0.20 & 49.15 & n.a.\\
Simul.$^\dagger$ & 0 & 127$\cdot10^7$ & 395 & 1343.49 & n.a.\\
\hline
\hline
Simul.$^*$ & 0 & 31 $\cdot 10^6$ & 1377 & 1078.57 & n.a.\\
\hline
\multicolumn{6}{l}{$^\dagger$ Results collected on Apple M1 Pro}\\
\multicolumn{6}{l}{$^*$ Results scaled from those on Apple M1 Pro}\\
\end{tabular}
}
\vspace{-0.4cm}
\end{table}

The results demonstrate the flexibility of a data-driven approach: we obtain configurations with latency, energy, and memory values that vary approximately by a factor of 3 (e.g., from 25ms/0.94$\mu$J to 68ms/2.6$\mu$J per regression and from 13.1 to 39.5 kB), with corresponding MAE values ranging from 7.2 to 14\%. Considering that we use a window of 2 hours as input to the models, the energy consumption values obtained by the data-driven solutions can be considered completely negligible.

Furthermore, on the M1 Pro CPU, our lowest error model pays a 7.2\% MAE in exchange for a striking 2,000X reduction in latency and 27X reduction in memory, with respect to directly executing a simulation-based SOH model. The latter, when scaled to the RISC-V platform, would require more than 1s to execute, implying a much higher energy overhead and more than 1MB of total memory, which would exceed the available space on most embedded microcontrollers.

\section{Conclusions}\label{sec:concl}
Data-driven models are the most suitable option for a digital twin of a battery to be hosted on-board a BMS, 
but their fidelity strongly depends on the quality of the training dataset.
We have shown that it is possible to use a simulation model to generate an arbitrarily large dataset in a much smaller time than that required by datasets obtained through measurements.
As results showed, this option also allows a tradeoff between model accuracy and model execution time or memory.

\bibliographystyle{IEEEtran}
\bibliography{biblio}

\end{document}